\documentclass[twocolumn,prb,showpacs,superscriptaddress,preprintnumbers,amsmath,amssymb]{revtex4}

\usepackage{graphicx}
\usepackage{dcolumn}
\usepackage{bm}

\begin{document}


\title{Crystal-field ground state of the non-centrosymmetric superconductor CePt$_3$Si:\\
a combined polarized soft X-ray absorption and polarized neutron study}

\author{T. Willers}
  \affiliation{Institute of Physics II, University of Cologne,
   Z{\"u}lpicher Stra{\ss}e 77, D-50937 Cologne, Germany}
\author{B. F{\aa}k}
\affiliation{Commissariat \`a l'Energie Atomique, INAC, SPSMS, 38054 Grenoble, France}
\author{N. Hollmann}
  \affiliation{Institute of Physics II, University of Cologne,
   Z{\"u}lpicher Stra{\ss}e 77, D-50937 Cologne, Germany}
\author{P. O. K\"orner}
  \affiliation{Institute of Physics II, University of Cologne,
   Z{\"u}lpicher Stra{\ss}e 77, D-50937 Cologne, Germany}
\author{Z. Hu}
 \affiliation{Institute of Physics II, University of Cologne,
   Z{\"u}lpicher Stra{\ss}e 77, D-50937 Cologne, Germany}   
\author{A. Tanaka}
  \affiliation{Department of Quantum Matter, ADSM Hiroshima
  University, Higashi-Hiroshima 739-8530, Japan}
\author{D. Schmitz}
\affiliation{Helmholtz-Zentrum Berlin, BESSY II, Albert-Einstein-Stra{\ss}e 15, D-12489 Berlin,
  Germany}
\author{M. Enderle}
  \affiliation{Institut Laue Langevin, 6 rue Horowitz, 38042 Grenoble, France}
\author{G. Lapertot}
  \affiliation{Commissariat \`a l'Energie Atomique, INAC, SPSMS, 38054 Grenoble, France}
\author{L. H. Tjeng}
  \affiliation{Institute of Physics II, University of Cologne,
   Z{\"u}lpicher Stra{\ss}e 77, D-50937 Cologne, Germany}
\author{A. Severing}
  \affiliation{Institute of Physics II, University of Cologne,
   Z{\"u}lpicher Stra{\ss}e 77, D-50937 Cologne, Germany}   

\date{\today}

\begin{abstract}
We determined the  crystal-field split Hund's rule ground
state of the non-centrosymmetric heavy fermion superconductor
CePt$_3$Si  with polarization dependent soft X-ray absorption
spectroscopy (XAS) and polarized neutron scattering. We are also
able to give the sequence of the crystal-field states from the 
temperature evolution of the linear dichroic signal in the XAS. The
quantitative analysis of the XAS temperature dependence together 
with the neutron transition energies complete the identification of 
the crystal-field level scheme.
\end{abstract}

\maketitle

\section{Introduction}

The discovery of superconductivity in the antiferromagnetic phases of the 
non-centrosymmetric heavy fermion compounds  
CePt$_3$Si,\cite{BauerPRL92} CeIrSi$_3$,\cite{Sugitani75} and 
CeRhSi$_3$ \cite{Kimura95} stimulated a flurry of theoretical and experimental 
activities \cite{BauerPSJ76, FujimotoPSJ76, MathurNature394, SamokhinPRB69} with the main 
interest to understand the pairing state in the absence of inversion symmetry. 
It seems that the superconductivity is due to a mixture of singlet and triplet Cooper
pairing and that the superconducting state is magnetically 
driven.\cite{GorkovPRL87, SaxenaNature427} The unusual properties evolve from 
the hybridized $4f$ electrons so that the question arises which influence the highly 
anisotropic crystal-field ground state may have on the hybridization between 
$4f$ and conduction electrons. Attempts were made to investigate this issue
for the heavy fermion compound CeRu$_2$Si$_2$,\cite{Zwicknagl1992} and the 
semiconducting Kondo materials CeNiSn,\cite{Ikeda1996, Kagan1997} and CeRhAs 
\cite{Ishi2004}. More recently the aspect of $q$-dependent hybridization was 
also studied in the CeMIn$_5$ compounds.\cite{Mena2005, Burch2007, Ghaemi2007, 
WeberPRB78} However, reliable experimental input for such theoretical 
investigation is scarce due to experimental uncertainties in determining 
the $4f$ ground state wave functions. 

The wave functions and crystal-field levels 
are usually determined from a combined analysis of single crystal susceptibility 
and polycrystalline neutron scattering measurements. 
However, the method often fails because of thermal or powder averaging in either technique,
and additionally, neutron scattering suffers from phonon contributions
and broadened crystal-field levels due to hybridization effects.\cite{Park1998}
Moreover, in the presence of magnetic order,
anisotropic molecular field parameters 
have to be introduced in order to fit the static susceptibility, ending 
up with too many free parameters for a unique description.  
The methods to apply are then inelastic or elastic polarized neutron 
scattering on single crystals, the first technique requiring large 
single crystals, the second being very sensitive to absorption 
corrections when determining the $4f$ magnetic form factor. 

We have shown for the case of the heavy fermion material CePd$_2$Si$_2$ 
 that soft X-ray absorption at the Ce M$_{4,5}$ edges 
can be used complementary to neutron scattering in order to determine the 
ground state wave function.\cite{HansmannPRL100} 
XAS is highly sensitive to the initial state
and via its  polarization dependence direct information about the 
$|J_z \rangle$ admixtures of the ground state wave function can be obtained. 
Sensitivity for higher lying crystal-field states is achieved by thermally 
populating those states. It was shown in Ref.~\onlinecite{HansmannPRL100} that the 
capability of XAS to resolve closely lying crystal-field states is not 
determined by the energy resolution of the experimental set-up, 
but rather by the temperature. However, the exact energy of 
crystal-field excitations is most accurately determined with inelastic
polarized neutron scattering on single crystals where phonon and magnetic 
scattering can be separated. Hence it is ideal to combine both techniques.  

CePt$_3$Si orders antiferromagnetically at T$_N$=2.2 K 
and enters a superconducting state at T$_{sc}\!\approx$0.75 K.\cite{BauerPRL92} 
The coexistence of both states has been confirmed by several experimental 
techniques.\cite{YogiPRL04,AmatoPRB05}
The low-energy magnetic excitations were recently measured
on a single crystal using inelastic neutron scattering.\cite{FakPRB78}
CePt$_3$Si crystallizes in the 
non-centrosymmetric tetragonal space group $P4mm$ with 
lattice constants $a$ = 4.072 {\AA} and $c$ = 5.442 {\AA}.\cite{Tursina04}
Inversion symmetry is broken by the absence of a mirror plane perpendicular to the 
$c$-axis. 
However, the lack of inversion symmetry is irrelevant for the crystal-field 
description since the $4f$ wave functions are even functions.
Therefore, as in other centrosymmetric compounds, 
the Hund's rule ground state of Ce$^{3+}$  
($J=5/2$, point group $C_{4v} \in D_{4h}$)\cite{FrigeriPRL92}
splits under the influence of a tetragonal crystal-field,
into three Kramer's doublets, which can be represented in the basis 
of $|J_z \rangle$. The eigenfunctions of the three Kramer's doublets 
can be written as
\begin{eqnarray}
|2 \rangle &=& \;\,\, |\pm1/2 \rangle\nonumber\\
|1 \rangle &=& \beta |\pm 5/2\rangle - \alpha |\mp 3/2\rangle\label{EqCFScheme}\\
|0 \rangle &=& \alpha |\pm 5/2\rangle + \beta |\mp 3/2 \rangle\nonumber
\end{eqnarray}
with $\alpha^2+\beta^2 = 1$ and an arbitrarily chosen phase of the 
mixed states. 

Two groups have published inelastic neutron 
scattering data of CePt$_3$Si powder samples in order to determine the 
crystal-field scheme. Bauer et al. \cite{BauerPB2005} found a broad 
distribution of magnetic intensity between 10~meV and 30~meV which 
was fitted with two non-resolved crystal-field excitations centered 
at 13~meV and 20~meV. The crystal-field parameters of Bauer et al. 
yield $|0 \rangle = 0.59 |\pm 5/2\rangle + 0.81 |\mp 3/2 \rangle$ 
for the ground state wave function and $|2 \rangle = |\pm1/2 \rangle$ 
as the highest excited state at 20~meV.  Metoki et al. \cite{MetokiJCM16} 
found also a mixed ground state, but interpreted their neutron data with a 
very different ground state mixing of $|0 \rangle = 0.93 |\pm 5/2\rangle 
+ 0.37 |\mp 3/2 \rangle$. According to their data the $|\pm1/2 \rangle$ 
state is 1~meV above the ground state and the second mixed state is at 
24~meV. 

In this paper, we determine the ground state wave function from the 
low temperature XAS data, i.e.\ the value of the mixing factor $\alpha$, 
which gives the actual spatial distribution of the crystal-field 
ground state.
We will further show that the temperature dependence of 
the XAS data determines the sequence of states. A quantitative modeling of 
the temperature dependence can even be done 
provided that the crystal-field energies are known. 
The latter are determined from the polarized inelastic neutron data 
presented here. Thus a complete picture of the crystal-field level 
scheme has been achieved. 

\section{Experimental}
The CePt$_3$Si single crystal used for this study was the same
as in Ref. ~\onlinecite{FakPRB78}.  A thin slice was cut from this crystal 
and used for the XAS measurements. We  recorded all
spectra using the total electron yield method in a chamber with a
pressure of 4~x~10$^{-11}$ mbar at the UE-46 PGM-1 undulator beam
line of BESSY II. The total electron yield signal was normalized
to the incoming photon flux $I_0$ as measured at the refocusing
mirror. Clean sample surfaces were obtained by cleaving the samples
{\it in situ}. The energy resolution at the cerium $M_{4,5}$ edges
($h\nu \approx 875-910$ eV) was set at 0.15 eV. Further, the
undulator together with a normal incident measurement geometry
allow for a change of polarization without changing the probed
spot on the sample surface. This guarantees a reliable
comparison of the spectral line shapes. The two polarizations 
were $E\!\perp\!c$ and $E\!\parallel\!c$, $c$ being the long tetragonal 
axis. The sample was cleaved and then cooled down to the lowest 
accessible temperature. Spectra were taken at 5~K, 50~K, 150~K, and 
200~K while warming up. In order to assure reproducibility of the data 
the sample was cleaved a second time and then data were taken in the 
sequence 300~K, 50~K and 80~K. 

Inelastic neutron scattering measurements using polarization analysis 
were performed on the thermal triple-axis spectrometer IN20 
at the high-flux reactor of the Institut Laue-Langevin. 
Horizontally focusing Heusler monochromator and analyzer were used 
in combination with relaxed collimation 
and a fixed final energy of 14.7 meV with an energy resolution 
(full width at half maximum, FWHM) of 0.6 meV.
Higher-order neutrons were removed from the scattered beam using a graphite filter 
and their contribution to the monitor count rate was corrected for. 
Measurements were performed in the paramagnetic phase at a temperature of 5 K 
with the wave-vector transfer {\bf Q} either parallel to the $a$ or to the $c$ axis, 
but with the same value of $Q=|{\bf Q}|$, 
to eliminate the influence of the magnetic form factor to the measured intensities. 
The use of polarization analysis in combination with a single crystal 
has two advantages for crystal-field determinations
compared with standard methods of unpolarized neutrons on a polycrystalline material: 
(i)~Scattering from phonons can easily be subtracted;
(ii)~Separation of the $j=x,y,z$ components of the angular momentum operator 
$M_j$ to the transition matrix elements $|M_j^\perp|^2\!\equiv\!|\langle f|M_j^\perp|i\rangle|^2$
between the initial state $i$ and the final state $f$, 
where $\perp$ reflects the fact that only components perpendicular to 
{\bf Q} are observed in neutron scattering experiments. 

We measured the three polarization components $\gamma=(\xi,\eta,\zeta)$ 
of the spin-flip cross-section $\sigma_\gamma$, 
where $\xi$ is parallel to {\bf Q},
$\eta$ is perpendicular to {\bf Q} in the horizontal scattering plane,
and $\zeta$ is vertical. The pure magnetic scattering is given by the 
difference of the spin-flip cross-sections $\!\parallel\!{\bf Q}$ and $\!\perp\!\bf{Q}$ 
or, more specifically for the present set-up, the different transition matrix elements 
of the $M_j$ operator are given by
$|M_z|^2\!=\!\sigma_\xi-\sigma_\eta$ for ${\bf Q}\!\parallel\!{\bf a}$ and 
$|M_x|^2\!=\!|M_y|^2\!=\!\sigma_\xi-\sigma_\zeta$ for ${\bf Q}\!\parallel\!{\bf a}$ or 
$|M_x|^2\!=\!|M_y|^2\!=\!\sigma_\xi-\sigma_\eta=\sigma_\xi-\sigma_\zeta$ 
for ${\bf Q}\!\parallel\!{\bf c}$. The three independent determinations 
of $|M_x|^2$ are identical within statistical uncertainty. 
The measured non-spin-flip cross-section shows no strong intensity 
that could leak through to the spin-flip channels via the finite flipping 
ratio of 20--23, which corresponds to a neutron polarization efficiency 
of 95\% at both polarizer and analyzer.

\section{Results and discussion}
For the C$_{4v}$ point symmetry the crystal-field Hamiltonian 
$H_{CF}=B_2^0 O_2^0 + B_4^0 O_4^0 + B_4^4 O_4^4$ describes the 
crystal field potential when the three Stevens parameters $B_2^0$,
$B_4^0$, and $B_4^4$ are determined. Alternatively the crystal-field
potential can be expressed in terms of the transition energies 
$E_1$ and $E_2$ between the crystal-field 
states within the multiplet and the mixing parameter $\alpha$.
We performed ionic full multiplet calculations using
the XTLS 8.3 program\cite{TanakaJPSC63} to calculate the
XAS spectra. All atomic parameters were given by Hartree-Fock values,
with a reduction of about 35\% for the $4f-4f$ Coulomb interactions
and about 20\% for the $3d-4f$ interactions to reproduce best the
experimental isotropic spectra, 
$I_{\rm iso} = 2 I_{\perp} + I_{\parallel}$. This accounts for the 
configuration interaction effects not included in the Hartree-Fock scheme. 
Once the atomic parameters are fine tuned to the isotropic spectra, 
the polarized XAS data can be described by the incoherent sums of the 
respective polarization dependent spectra of the pure $|J_z \rangle$ 
states \cite{HansmannPRL100} as long as the crystal-field splitting 
is small with respect to the spin orbit splitting. The latter requirement, 
which is fulfilled here ($E_{SO}$ $\approx$ 280~meV and $E_{CF}$ 
$\leq$ 20~meV) assures that interference terms resulting 
from intermixing between the $J=5/2$ and $J=7/2$ multiplet can 
be neglected. 

\begin{figure}[]
    \centering
    \includegraphics[width=1.0\columnwidth]{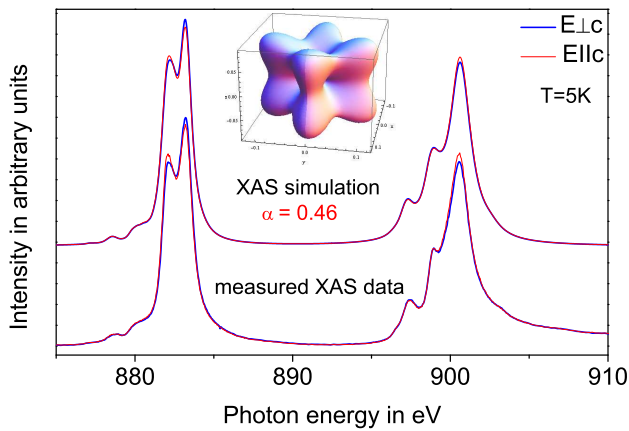}
    \caption{(color online) Measured and simulated linear polarized XAS spectra of CePt$_3$Si at 
    the M$_{4,5}$ edges at $T$=5 K. The inset shows the corresponding spatial distribution.}
    \label{fig:5K}
\end{figure}

Figure 1 shows the low temperature linear polarized XAS data of CePt$_3$Si 
at the M$_{4,5}$ edge and their simulation based on the full multiplet 
treatment as described above. The polarization effect is not very 
large but real as will be shown in the next paragraph from the temperature 
dependence of the XAS data and the reproducibility of the 50~K spectra 
after the second cleave. The pure $ |\pm 1/2 \rangle$ state cannot
describe our low temperature data as can be seen from Fig.~2, where we 
have shown again (see Ref.~\onlinecite{HansmannPRL100}) for reasons of 
convenience, the polarization dependent spectra for each $|J_z \rangle$ state. 
The 5~K data are well described with the mixed ground state
\begin{equation*}
|0 \rangle = 0.46 |\pm 5/2\rangle + 0.89 |\mp 3/2 \rangle\nonumber
\end{equation*}
($\alpha = 0.46 \pm 0.01$) without the necessity to include the fractional 
occupation of a higher lying state. The small size of the polarization effect 
can be understood when considering, as was shown in the paper by Hansmann 
et al.,\cite{HansmannPRL100} that the variation of $\alpha$ from 0 to 1 
leads to a vanishing polarization effect when $\alpha$ is equal to 
$\sqrt{1/6}$ = 0.41, i.e.\ when $\alpha$ is equal to the value of a cubic 
crystal-field symmetry. According to our data the mixing factor in CePt$_3$Si 
is close to this value. 

\begin{figure}[]
    \includegraphics[width=1.0\columnwidth]{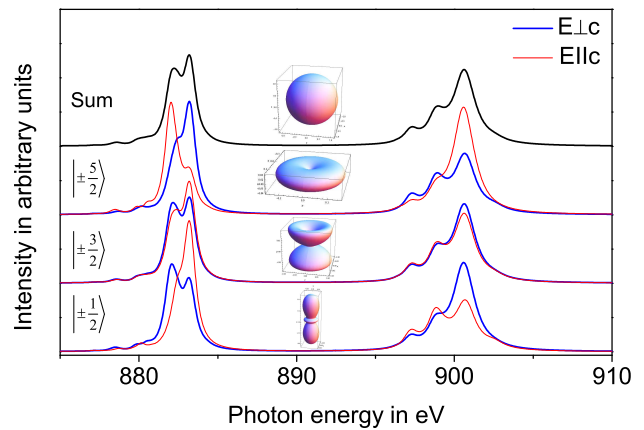}
    \caption{(color online) Simulated polarizations for pure $|J_z \rangle$ states of Ce$^{3+}$ 
    and sum of the three polarization dependent $|J_z \rangle$ spectra. 
    The insets resemble the spatial distributions.}
    \label{fig:States}
\end{figure}

\begin{figure*}
    \centering
    \includegraphics[width=2.0\columnwidth]{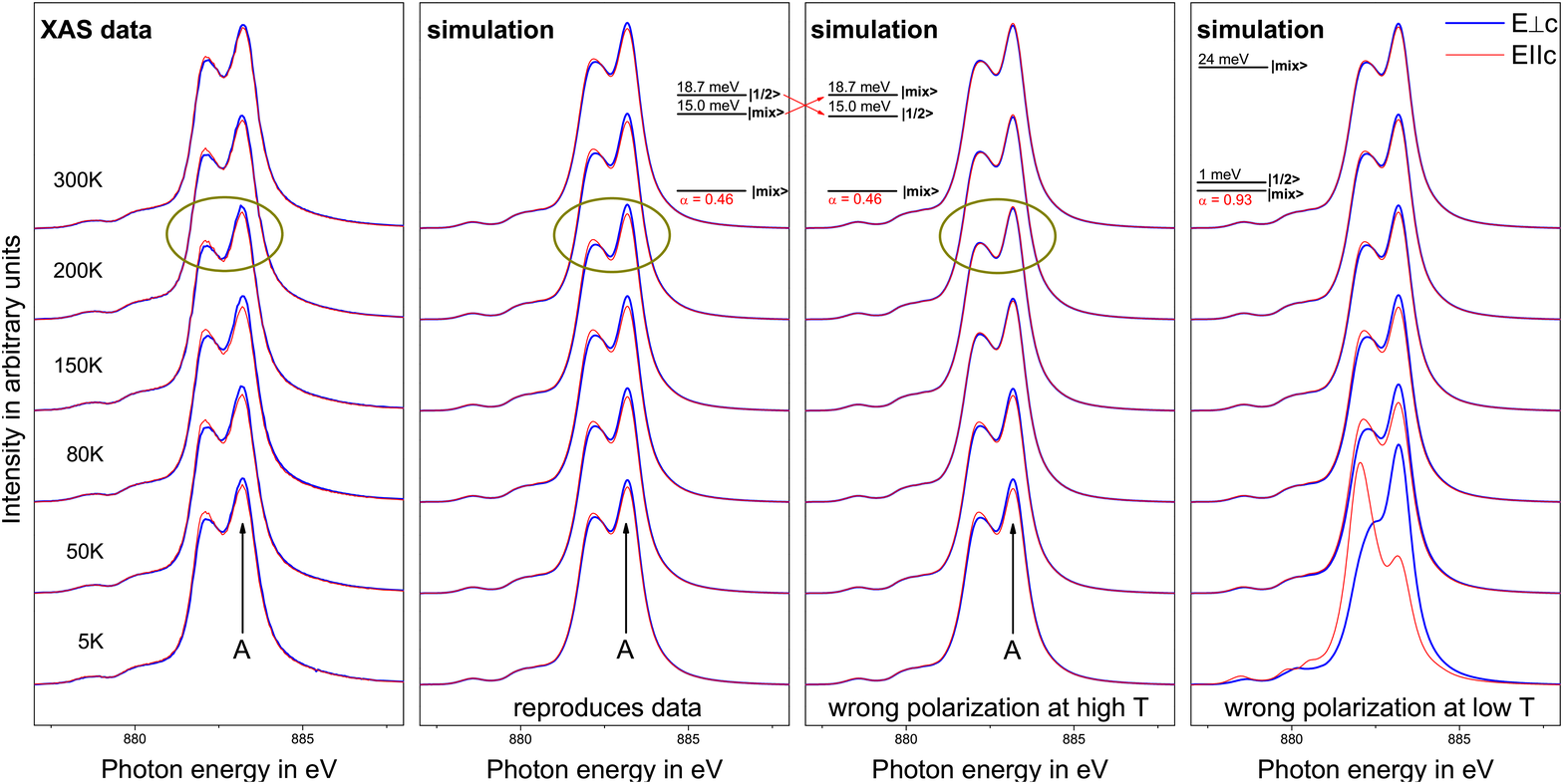}
    \caption{(color online) Left panel: temperature dependent XAS data at the M$_5$ edge. Two middle 
    panels: simulation of spectra with ground state wave function as determined from the 
    5 K XAS data and with crystal field energies obtained from the polarized neutron 
    data. The simulations in the middle columns have different sequences of states. Right panel:
    simulation with ground state wave function and energies as given by Metoki 
    et al.\cite{MetokiJCM16}}
    \label{fig:3Temp}
\end{figure*}

In order to rule out that the small size of the effect is due to unwanted 
surface effects the sample was measured at several temperatures, cleaved 
a second time and remeasured. The left panel side of Fig.~3 shows the 
M$_5$ edge of CePt$_3$Si for several temperatures between 5~K and 300~K. 
After the first cleave data were taken at 5~K, 50~K, 150~K and 200~K. 
After the second cleave the sample was measured first at 300~K and then 
cooled down to 50~K and warmed up to 80~K. Comparing the data at different 
temperatures shows that the polarization increases from 5~K to 50~K and 80~K 
and then decreases again as temperature rises further. We 
conclude that this effect is real since the polarization at 300~K 
(measured first after second cleave) is smaller than the polarization at 
50~K (measured second after second cleave) and 80~K (measured third after 
second cleave). Surely, if surface degradation with time had been present, then 
one would rather expect to see a decrease of polarization.
In addition the 50~K data were well reproduced. The black dots in Fig.~4 are 
the difference of the two intensities $E\!\perp\!c$ and $E\!\parallel\!c$ 
(called \textit{linear dichroism} LD) at the high energy peak $A$ of the M$_5$ edge 
(see Fig. 3) and they summarize the temperature evolution of the polarization effect. 

\begin{figure}
    \centering
    \includegraphics[width=1.0\columnwidth]{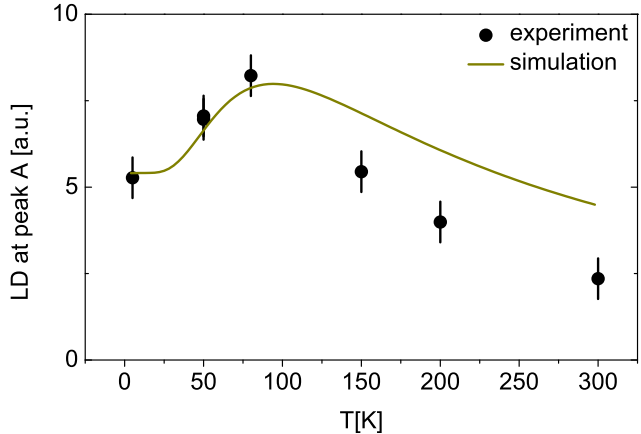}
    \caption{(color online) Difference of the intensities $E\!\perp\!c$ and $E\!\parallel\!c$
    (\textit{linear dichroism} LD) at peak $A$ of the M$_5$ edge 
    versus temperature. The black dots are determined from the experimental data 
    (left panel in Fig.~3), the dark yellow line results from a simulation 
    according to the left, middle panel in Fig.~3.}
    \label{fig:PversusT}
\end{figure}

Polarized inelastic neutron scattering was used to determine
the energy of the crystal-field levels. 
Figure 5 shows energy scans for {\bf Q}=(1.5,0,0) 
for difference cross-sections sensitive to $M_x$ and $M_z$, respectively. 
Within the crystal-field scheme of Eq.~(\ref{EqCFScheme}), 
the $M_z$ cross-section (open blue symbols in Fig.~5) contributes only to the transition
between the ground state $|0\rangle$ and the first excited state $|1\rangle$. 
The data were fitted by a damped harmonic oscillator (DHO)\cite{DHO}
convoluted with the instrumental resolution function.
The first excited state $|1\rangle$ has an energy 
(with respect to the ground-state energy) of $E_1$=15.0(4) meV
and a width (FWHM) of 8.2(9) meV, 
which reflects the damping of the crystal-field excitation
due to the hybridization with the conduction electrons. 
The $M_x$ cross-section
contains both transitions, $|0\rangle\rightarrow |1\rangle$ and 
$|0\rangle\rightarrow |2\rangle$,
and the corresponding data (full red symbols in Fig.~5) 
were modeled with two DHOs. 
The energy and width of the first transition (DHO) in the $M_x$ data
were kept fixed to the above values
while the integrated intensity was fixed to the fitted value of the $M_z$ cross-section
after scaling with the corresponding transition probabilities,
calculated from the mixing parameter $\alpha$ of the XAS measurement. 
The second DHO described the second transition, 
for which the best fit 
gave an energy of $E_2$=18.7(5) meV and FWHM = 6.2(8) meV. 
The fitted integrated intensity of the second DHO is about 16\% smaller 
than expected from the calculated transition probabilities. 
We consider this to be an excellent agreement between 
neutron intensities and our findings from XAS.

\begin{figure}
\centering
\includegraphics[width=\columnwidth]{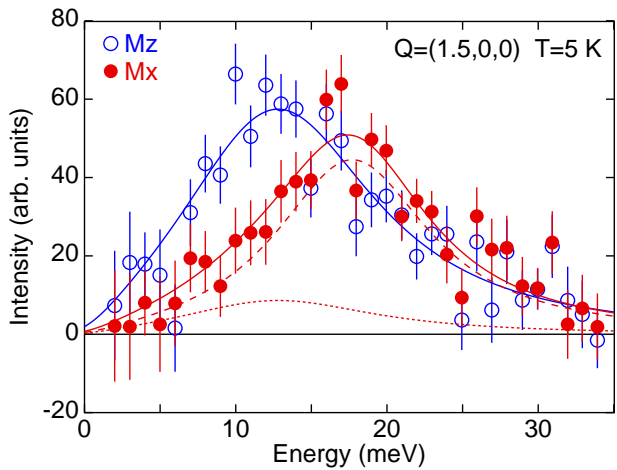}
\caption{(color online) Magnetic intensities as obtained from neutron scattering difference 
spectra taken at {\bf Q}=(1.5,0,0) and $T$=5 K. 
Open blue symbols and line: experimental data of the $|M_z|^2$ component 
and fit to the transition $|0\rangle\rightarrow |1\rangle$. 
Full red symbols: experimental data of the $|M_x|^2$ component, 
which includes the transition $|0\rangle\rightarrow |1\rangle$
(dotted red line, all parameters fixed, see text)
as well as the transition $|0\rangle\rightarrow |2\rangle$
(dashed red line, all parameters free). 
The full red line is the sum of both contributions.}
\label{Figure5}
\end{figure}

The temperature dependence of the XAS data can now be modeled quantitatively using
the ground state wave function as determined from XAS at $T$= 5 K and the energy 
levels as determined with polarized neutrons. 
The results of these calculations are
shown in the two middle panels of Fig.~3. The simulations differ in the order
of the first and second excited state. Only the simulation with the 
$|\pm 1/2\rangle$ as highest state reproduces the temperature dependence of 
the data well. The other option with the $|\pm 1/2\rangle$ as first excited 
state leads to an inversion of polarization as temperature rises: for example, the 
intensity of peak $A$ for $E\!\perp\!c$ (blue) is simulated to become smaller than 
for $E\!\parallel\!c$ (red) at 150 K, in contradiction with the experiment, see dark yellow circles
in the left and middle right panel of Fig.~3. This can be
understood when looking at the pure states as displayed in Fig.~2 and keeping in mind
that the XAS spectra at finite temperatures resemble the superimposed polarizations
of each state. The top of Fig. 2 shows that the full thermal population of 
all crystal-field states leads to a vanishing polarization since the sum of all 
$|J_z \rangle$ spatial distributions is spherical. From this 
can be concluded that a small ground state polarization requires the first and 
second excited state to almost compensate each other, i.e. the other mixed state 
must have a polarization almost opposite to the one of the $|\pm 1/2\rangle$ 
(see Fig.~2). Hence mixing in the other mixed state as temperature rises 
leads at first to an increase of polarization until the influence of the 
highest state, here the $|\pm 1/2\rangle$, compensates this due to thermal 
population. The dark yellow line in Fig.~4 displays the simulated temperature 
dependence of the polarization. The simulation reproduces the experimental 
temperature dependence quite well, including the presence of a 
maximum at 80~K.

For completeness we have simulated the XAS spectra using the 
crystal-field model suggested by Metoki et al.\cite{MetokiJCM16} 
Their ground state wave function produces spectra which do not 
resemble the experimental ones (see right panel of Fig.~3). Moreover, 
their model would lead to a strong temperature dependence between 
5~K and 50~K which is not present in the experimental data.

\section{Conclusion}
We have performed temperature dependent linear polarized XAS measurements 
at the M$_{4,5}$ edge of cerium in CePt$_3$Si and inelastic polarized 
neutron experiments. The combination of both techniques gives a conclusive 
picture of the crystal-field level scheme in CePt$_3$Si. The low temperature 
XAS data yield $0.46 |\pm 5/2\rangle + 0.89 |\mp 3/2 \rangle$ as the ground 
state wave function and the sequence of excited states can clearly be 
determined from the temperature dependence of the XAS data. Using 
the crystal-field transition energies as determined from the polarized 
neutron scattering data (15.0 and 18.7 meV) we were also able to quantitatively 
explain the temperature dependence. The wave-vector and polarization 
dependence of the neutron scattering intensities
also confirm this sequence of the crystal-field levels. 
We note that the result with having the $|\pm 1/2\rangle$ level as the 
highest lying excited state is in excellent agreement with 
the findings by Bauer et al.\ and excludes the suggestions by Metoki at al.

\section*{Acknowledgements}
The experiments at BESSY were supported by the BMBF through
project 05 ES3XBA/5. We thank L. Hamdan and the Cologne Mechanical 
Workshop for skillful technical support. The wave function 
density plots were calculated using the CrystalFieldTheory 
package for Mathematica written by M. W. Haverkort 
and the neutron transition matrix elements by a program written by
J.X. Boucherle and F. Givord.



\end{document}